\documentstyle[12pt,epsfig]{article}                                
\begin{document}

\titlepage                  

\begin{center}
\begin{Large}
\begin{bf}

Real Spectra in Logarithmic model PT-symmetry operators: Iso-spectra in Logarithmic PT-symmetry

\vspace{1.0cm}

\end{bf}
\end{Large}

 Biswanath Rath $\dagger$,Rabab Jarrar$*$, Hussein Shanak$88$, Jihad Asad$***$ and Rania Wannan$\bullet$

\end{center}

\vspace{0.1cm}

\begin{it}
$\dagger$ Department of Physics,
 Maharaja Sriram Chandra Bhanj Deo University,
 Takatpur, Baripada -757003, Odisha, India.
e.mail:biswanathrath10@gmail.com

\vspace{0.1cm}

$*;**;***$: Department of Physics,Faculty of Applied Sciences, Palestine Technical University, Kadoorie ,Tulkarm P 305,Palestine.
e.mail:j.asad@ptuk.edu.ps ;r.jarrar@ptuk.edu.ps; h.shanak@ptuk.edu.ps 

\vspace{0.1cm}

$\bullet$ Department of Applied Mathematics,faculty of Applied Sciences, Palestine Technical University, Kadoorie,Tulkarm,P 305, Palestine.
e.mail:r.wannan@ptuk.edu.ps

\vspace{0.1cm}

\end{it}

$\bf{Abstract:}$
We reflect real spectra of new logarithmic model PT-symmetry operators with singular and non-singular in nature. We also notice the iso-spectral nature between inverted and non-inverted PT-symmetry potentials. Present numerical result give good agreement with previous results.
with available  results.
\vspace{1.0cm}

\begin{bf}
\hspace{3.0in} PACS:11.30.Pb;03.65.Ge.
\end{bf}

\vspace{0.1cm}

\hspace{0.5cm} \noindent\rule{3.0in}{0.4pt}

Correspondence: biswanathrath10@gmail.com ; m.kanan@ubt.ps ; j.asad@ptuk.edu.ps 

\begin{bf}
I.Introduction
\end{bf}

Real spectra in quantum operators are confined to Hermiticity($H=H^{\dagger}$) as well as PT-symmetry [1]($[H,PT]=0$).Here $P$ stands for parity operator having the properties: $PxP^{-1}=-x$; $P p P^{-1} = -p$. Similarly $T$ stands for the time reversal operator having the properties$T xT^{-1}=x$;$T p T^{-1}=-p$ and $T  T^{-1}=-i$. In the Hermiticity(more precisely self-adjoint operator), it ia possible to find two Hamiltonians, which are iso-spectral to each other[2]. For 
example[2,3]  
\begin{equation}
h^{(1)}=p^{2} + V_{0}(1-e^{2|x|/a})
\end{equation}
and
\begin{equation}
h^{(2)}=p^{2} + v_{0}(1-e^{-2|x|/a})
\end{equation}
If one clearly analyzes one is scattering in nature and the other is confining nature [2,3].

Till now no such models are reflected in nature. In a recent paper Bender etal
 [4] have suggested a new class of logarithmic PT-symmetry potentials as 

\begin{equation}
h_{1}=p^{2} + x^{4} \log(ix)
\end{equation}
\begin{equation}
h_{2}=p^{2} - x^{4} \log(ix)
\end{equation}
\begin{equation}
h_{3}=p^{2} - x^{4} \log(x^{2})
\end{equation}

and reflected energy spectrum of $H_{1}$ only. Further authors  reported analytical calculation of energy level using WKB approach[4] 
\begin{equation}
\frac{E_{n}}{[\log{E_{n})}]^{1/3}} \sim \Biggl[ \frac{\Gamma(7/4)(n+1/2)\sqrt{\pi}}{\Gamma(5/4)\sqrt{2}}\Biggr]^{4/3}
\end{equation}
does not give encouraging results when compared with numerical results. This
 motivates the present author to calculate spectra of $H_{1,2}$ and suggest new models on logarithmic potentials. Apart from this aim is to find out whether 
iso-spectral Hamiltonians  are possible in PT-symmetry operators.

\begin{bf}
2.Logarithmic new models  
\end{bf}

Here we consider different models as follows

\begin{bf}
Quadratic Logarithmic  model $V(x)=-x^{2} \log(\frac{i}{x})$
\end{bf}

The Hamiltonian considered here is 
\begin{equation}
H_{1}^{quadratic}= p^{2} - x^{2} \log(\frac{i}{x})
\end{equation}

\begin{bf}
Quartic inverted model $V(x)=-x^{4} \log(\frac{i}{x})$
\end{bf}

Here we consider the Hamiltonian 
\begin{equation}
H_{2}^{quartic}= p^{2} - x^{4} \log(\frac{i}{x})
\end{equation}

Below we present few energy levels

\begin{table}[htbp]
\begin{center}

\vspace{1.0cm}

TABLE.I: PT-symmetry inverted logarithmic model potentials 

\begin{tabular}{ c c  c } \\ \hline

n & $H_{1}$ Present   & $H_{2}$ \\  \hline \hline  
0 & 1.326 591 6  & 1.249 087 3      \\
3 & 9.173 294 3 & 13.738 280 4    \\
6 & 17.734 002 2 &31.665 810 8   \\
9 & 26.633 530 1 & 52.993 926 2 \\  
12& 76.113 329 2 & 76.974 762 5 \\   \hline \hline  
\end{tabular} 
\end{center}
\end{table}

\begin{bf}
3.Logarithmic models  
\end{bf}

Here we consider the recently prposed models 
 
\begin{equation}
h_{1}=p^{2} + \lambda x^{4} \log(ix)
\end{equation}
\begin{equation}
h_{2}=p^{2} - \lambda x^{4} \log(ix)
\end{equation}
\begin{equation}
h_{3}=p^{2} -\lambda  x^{4} \log(x^{2})
\end{equation}

and present complete spectra in table-2 for $\lambda=1$.

\begin{table}[htbp]
\begin{center}

\vspace{1.0cm}

TABLE.2: PT-symmetry logarithmic model

\begin{tabular}{ c c  c c  } \\ \hline

n & $h_{1}$ Present   & $h_{1}$Previous [4] & Previous (WKB)[4] \\  \hline \hline  
0 & 1.249 08  & 1.249 09   & 0.546 27      \\
3 & 13.738 27 &13.738 3    &  7.314 80 \\ 
6 & 31.665 82 &31.665 8     & 16.697 9   \\
9 & 52.993 79 & 52.993 9  & 27.695 6 \\ 
12& 76.976 08 & 76.974 8  & 39.932 4 \\ \hline
n & $h_{2}$ Present & Previous[4] & Previous(WKB)[4]  \\  \hline
0 & 0.109 1 &  \\
1 & 6,959 6 &  \\ 
2 & 8.257 1 &  \\
3 & 18.039 4 & \\    \hline \hline 
n & $h_{3}$ Present   &Previous[4] & Previous(WKB)[4] \\  \hline 
0 & 0.025 4 &  \\
1 & 4.977 7 &    \\
2 & 9.237 1 & \\
3 & 16.478 6 &  \\ \hline \hline 
\end{tabular} 
\end{center}
\end{table}

\pagebreak 

\begin{bf}
4.Method of calculation
\end{bf}

Here we use matrix diagonalisation method [5] to solve the eigenvalue relation 
\begin{equation}
H|\Psi>= E|\Psi>
\end{equation}
where
\begin{equation}
|\Psi>=\sum A_{m}|m>
\end{equation}
Here $|m>$ satisfy the relation 
\begin{equation}
[p^{2}+x^{2}]|m>=(2m+1)|m>
\end{equation}
Numerical results obtained using MDM are tabulated in table.1.

\begin{bf}
5.Conclusion
\end{bf}

In this report, we present numerical convergent energy levels of new model PT-invariant Hamltonians using matrix diagonalisation method[5]. Further we feel the
  present method can be used confidently to realize real spectra study in 
similar Hamiltonians of interest. Lastly we  the spectra of
 $H_{2}$ and $h_{1}$ are the same.
In brief 
\begin{equation}
V(x) = x^{4}\log(ix) \rightarrow V(x) = - x^{4} \log(\frac{i}{x})
\end{equation}

 Hence these two potentials can be considered as  iso-spectral  models in PT-symmetry. Lastly we do not find any numerical results to present in table-1 for a comparison with the present numericals. Interested readers can consider other values of $\lambda$.

\begin{bf}
Authors contribution:
\end{bf}

B.Rath: formulation,computation,writing,finalizing;R.Jarrar: Computtation, finance;H.Shanak: computation and finance; J.Asad: writing,computation and Finance ; R.Wannan-computation,finance.

\begin{bf}
Conflict of interest
\end{bf}

Authors declare there is no conflict of interest.

\begin{bf}
DATA AVAILABILITY
\end{bf}

No additional data is required . All the datas  included in this paper are 
sufficient.

\begin{bf}
Declaration
\end{bf}

Present paper is a modified version of arxiv paper.


\begin{thebibliography}{99}

\bibitem{Bender}C.Bender and S.Boettcher Real spectra in non-Hermitian Hamiltonian having PT-symmetry, Phys.Rev.Lett,(1997),$\bf{80}$,5243-5246.
\bibitem{Jones}H.F.Jones,Comment on: Solvable model bound states in the continuum (BIC)in One dimension(2019, 94, 105214),Phys.Scr,(2021),$\bf{96}$,087001.(see ref-2)
\bibitem{Jones}Z.Ahmed and H.F.Jones,Scattering states and bound states of exponential potentials, arxiv:2102:06095v1(see ref-3).
\bibitem{Bender}C.Bender,A.Felski,S.P.Klevansky and S.Sarkar,PT-symmetry and Renormalisation in Quantum Field Theory,arxiv:2103.14864v1.
\bibitem{Rath}B.Rath,Real spectra in some negative potentials: Linear and nonlinear one dimensional PT-invariant quantum systems,Eur.Phys.Journal.Plus.(2021), $\bf{136}$,493.
\end{thebibliography}
\end{document}